\documentclass[12pt]{iopart}

\def\newblock{\hskip .11em plus .33em minus .07em}
\usepackage[pdftex]{graphicx}
\usepackage[sort&compress]{natbib}
\bibpunct{[}{]}{,}{n}{}{;}
\usepackage{url}

\newcommand{\betwo}[2]{\ensuremath{B(\mathrm{E2}:#1\to#2})}
\newcommand{\isotope}[2]{\ensuremath{^{#1}\textrm{#2}}}

\begin{document}

\title[Nuclear deformations in the region of the $A=160$ r-process abundance peak]{Nuclear deformations in the region of the $A=160$ r-process abundance peak}

\author{P.-A. S\"{o}derstr\"{o}m}

\address{Department of Physics and Astronomy, Uppsala University, SE-75120 Uppsala, Sweden}
\ead{P-A.Soderstrom@physics.uu.se}
\begin{abstract}
In the abundance spectrum of r-process nuclei the most prominent features are the peaks that form
when the r-process flow passes through the closed neutron shells. However, there are also other
features in the abundance spectrum that can not be explained by shell effects, like the peak in the
region of the rare-earth nuclei around mass $A=160$. It has been argued that this peak is related to the deformation maximum of the neutron-rich isotopes.
Recently, both experimental and theoretical work has been carried out to study the deformation of
neutron-rich rare-earth nuclei and to search for the point of maximum deformation. This work has
focused on the nuclei around $^{170}$Dy in order to understand the evolution of collectivity
in the neutron shell with $82 < N < 126$. These investigations will be discussed in terms of the Harris parameters of the
Variable Moment of Inertia model.
Finally, we will discuss the future
possibilities to reach further into the neutron-rich rare-earth region at the new experimental facilities
using radioactive beams.
\end{abstract}


\section{Introduction\label{ss:theory_r}}

The two predominant astrophysical mechanisms for production of heavy elements in the universe are the slow neutron-capture process (s-process) and the rapid neutron-capture process (r-process) \cite{RevModPhys.29.547}. The r-process occurs in high neutron-density areas in the universe, where the average time for neutron capture is smaller than the half life of the radioactive isotope, and an equilibrium between neutron capture, (n,$\gamma$), and photodisintegration, ($\gamma$,n), has established itself. After this so called steady phase of the r-process, the free neutrons disappear and the nuclei $\beta$ decay back to stability, a process called freeze out. The most dominant features of the abundance distribution of elements created during the steady phase and freeze out are the large peaks at $A\approx80$, $A\approx130$ and $A\approx195$ that are due to the r-process flow through closed shells. The second most pronounced feature is the peak at $A\approx160$ in the rare-earth region.
While the closed shell peaks are well understood to be formed during the steady phase \cite{RevModPhys.29.547}, it has been argued that the $A\approx160$ peak is due to the deformation in the nuclei created after the steady phase, just before freeze out \cite{surman1}. An explanation that has been proposed is that as the deformation maximum is reached the nucleus cannot deform more so the next heavier nucleus will be less stable, an effect that can mimic closed shells.

\section{Deformation systematics}

\begin{figure}
 \centering
 \includegraphics[width=0.8\textwidth]{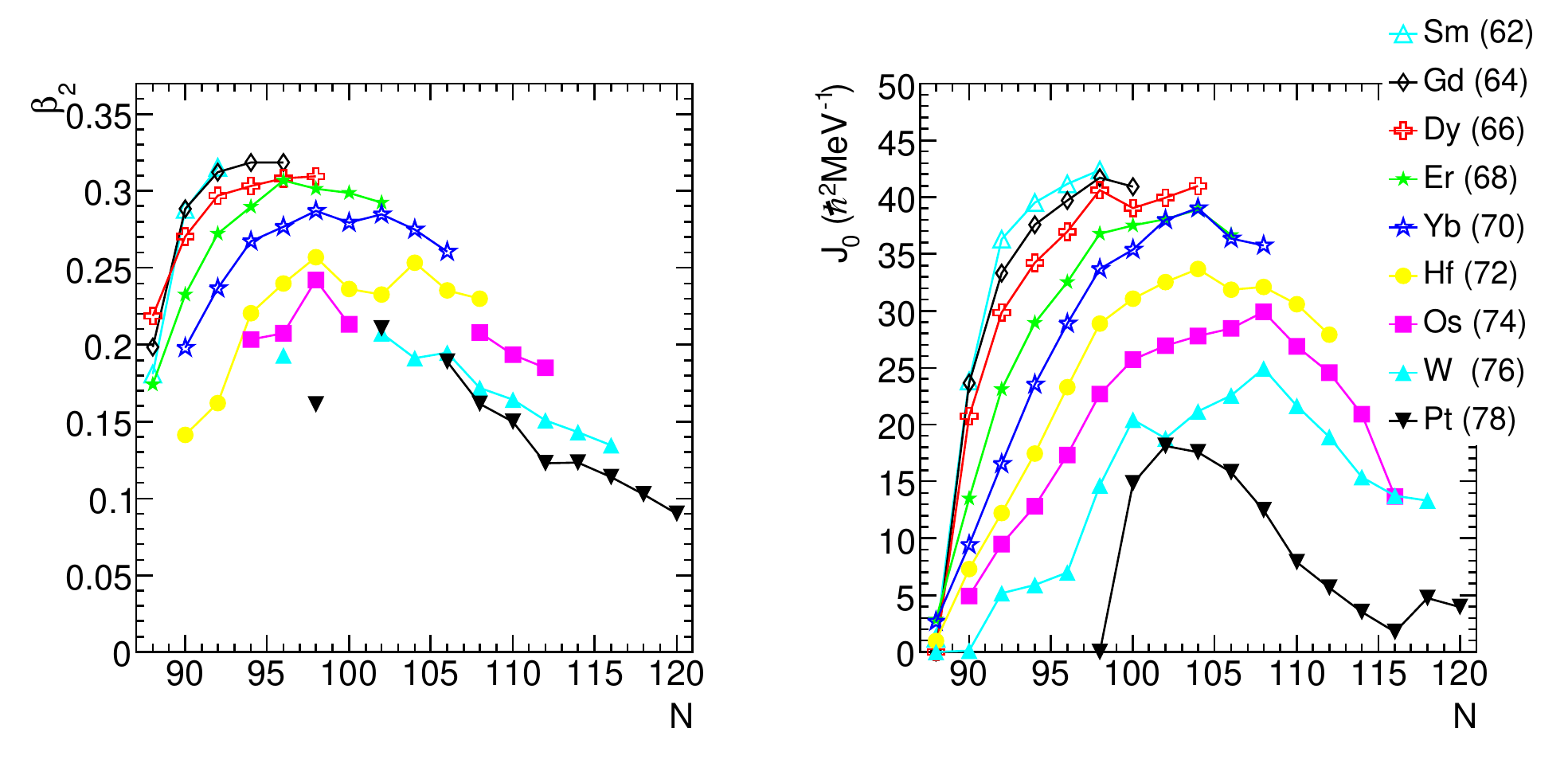}
 \caption{Experimentally measured $\beta_2$ values (left) and $J_0$ Harris parameters (right) for a selection of even-even nuclei \protect\cite{my_thesis}.}
 \label{fig:mnsystematics}
\end{figure}

One of the standard references for nuclear masses, and deformations, is the calculations made by M\"{o}ller and Nix using the finite range liquid drop model \cite{mollernix}. This reference has, for example, been used in the calculations in \cite{surman1, surman2}, reproducing the $A=160$ r-process peak position but slightly underestimating the low-$A$ side and slightly overestimating the high-$A$ side. In \cite{mollernix}, the deformations behave smoothly with a maximum in the region around $N\approx102$--$104$. In figure~\ref{fig:mnsystematics}, the evaluated experimental deformations from \cite{nndc} are shown, but the calculations from \cite{mollernix} do not appear to follow the same pattern as the experimental data. Besides the larger absolute variation in deformations for different $Z$ values, the deformation maximum for each $Z$ does not appear to be as stable around $N\approx104$ as in the M\"{o}ller and Nix calculations. Rather, the deformations seem to peak at lower values of $N$ for lower values of $Z$, further discussed in \cite{naturalVMI}.

\begin{figure}
 \centering
 \includegraphics[angle=-90,width=0.8\textwidth]{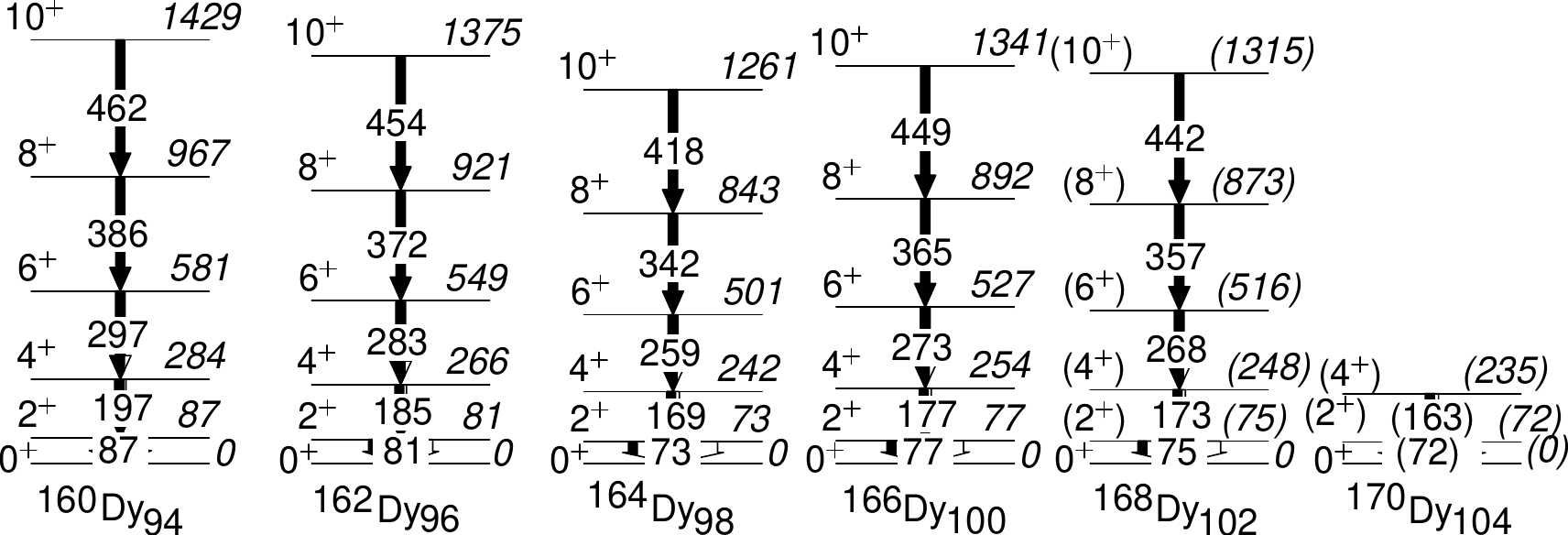}
\caption[Ground state rotational bands for dysprosium isotopes with $N=94-104$]{Ground state rotational bands for dysprosium isotopes with $N=94-104$ from \protect\cite{nndc} and for $6^+$--$10^+$ in $^{168}$Dy and the $4^+\to2^+$ transition in \isotope{170}{Dy} \protect\cite{PhysRevC.81.034310}. The $2^+\to0^+$ transition in \isotope{170}{Dy} is from the calculations in \protect\cite{chinphyslett}.}
 \label{fig:dy_syst}
\end{figure}

As the experimental data on nuclear deformation is quite sparse in this neutron-rich region, other ways of understanding the evolution of nuclear deformations have to be investigated. For example, one can use the excitation energy spectrum together with the VMI model, which is based on the Harris parameters $[J_0,J_1]$. The parameter $J_0$ can be related to the deformation of the nucleus and $J_1$ can be related to the amount of freezing of the internal structure, that is the rigidity. In this way it is possible to obtain a much richer set of data than when using only experimental deformations directly. Recently, an experiment to study the yrast band in the mid-shell nuclei \isotope{168,170}{Dy} ($N=102,104)$ was carried out at the PRISMA and CLARA set-up at LNL, the results of which are shown in figure~\ref{fig:dy_syst}. The VMI fits have been made according to the procedure in \cite{naturalVMI}, with the inclusion of the data on \isotope{168}{Dy} \cite{PhysRevC.81.034310} and \isotope{170}{Dy} \cite{PhysRevC.81.034310,chinphyslett}. The experimental data from \cite{nndc} is shown together with the Harris parameters $J_0$ in figure~\ref{fig:mnsystematics}. 
These new data points show an increase in $J_0$, suggesting that the $N=104$ deformation maximum could be reasonably stable at least from Dy to Hf, $66\leq Z \leq 72$.
However, further investigations at lower $Z$ and higher $N$, as well as direct measurements of the deformation parameters, are important for a full understanding of the evolution of nuclear deformations.

\section{Outlook}

\begin{figure}
 \centering
 \includegraphics[width=0.4\textwidth]{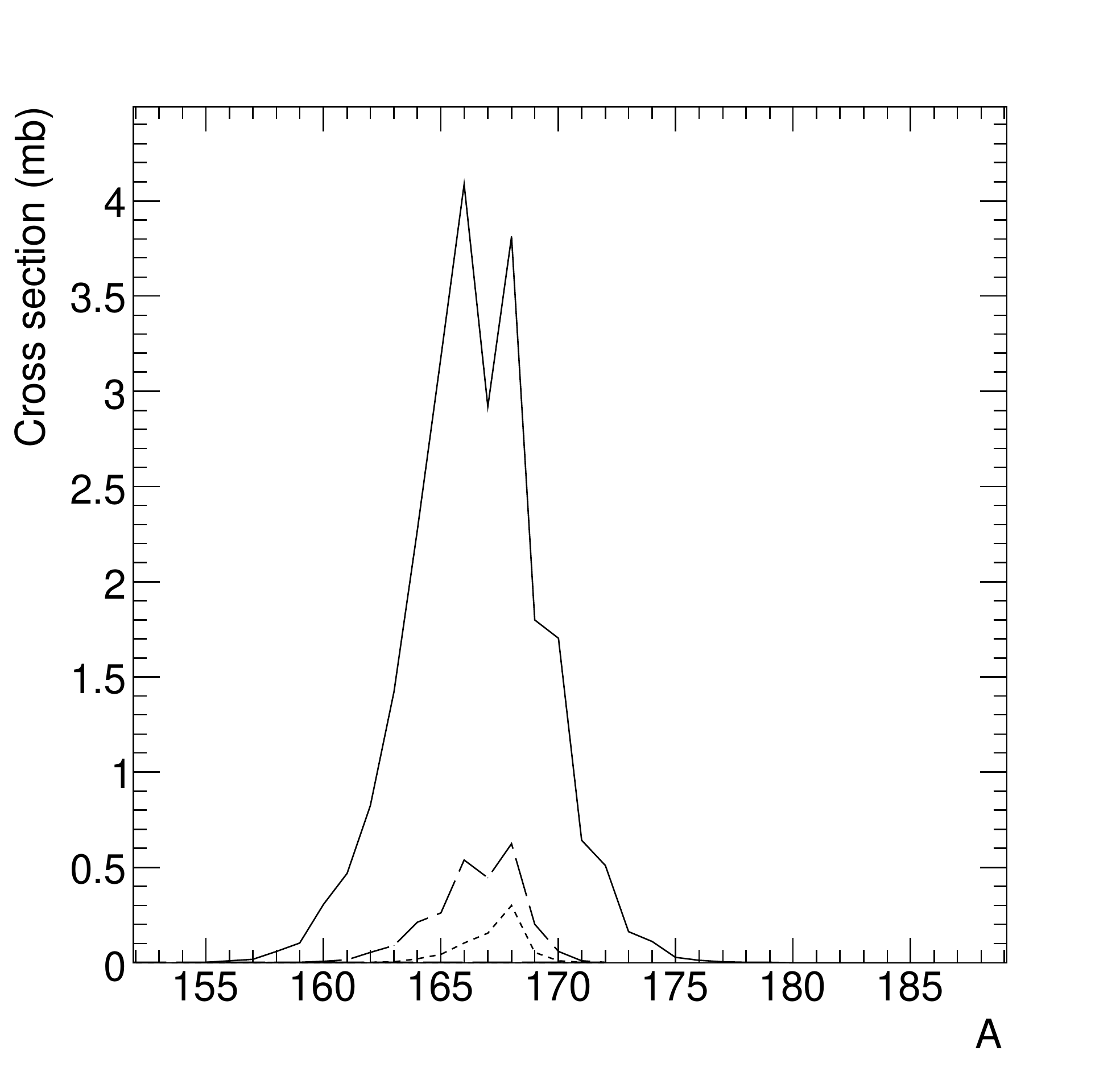}
 \includegraphics[width=0.4\textwidth]{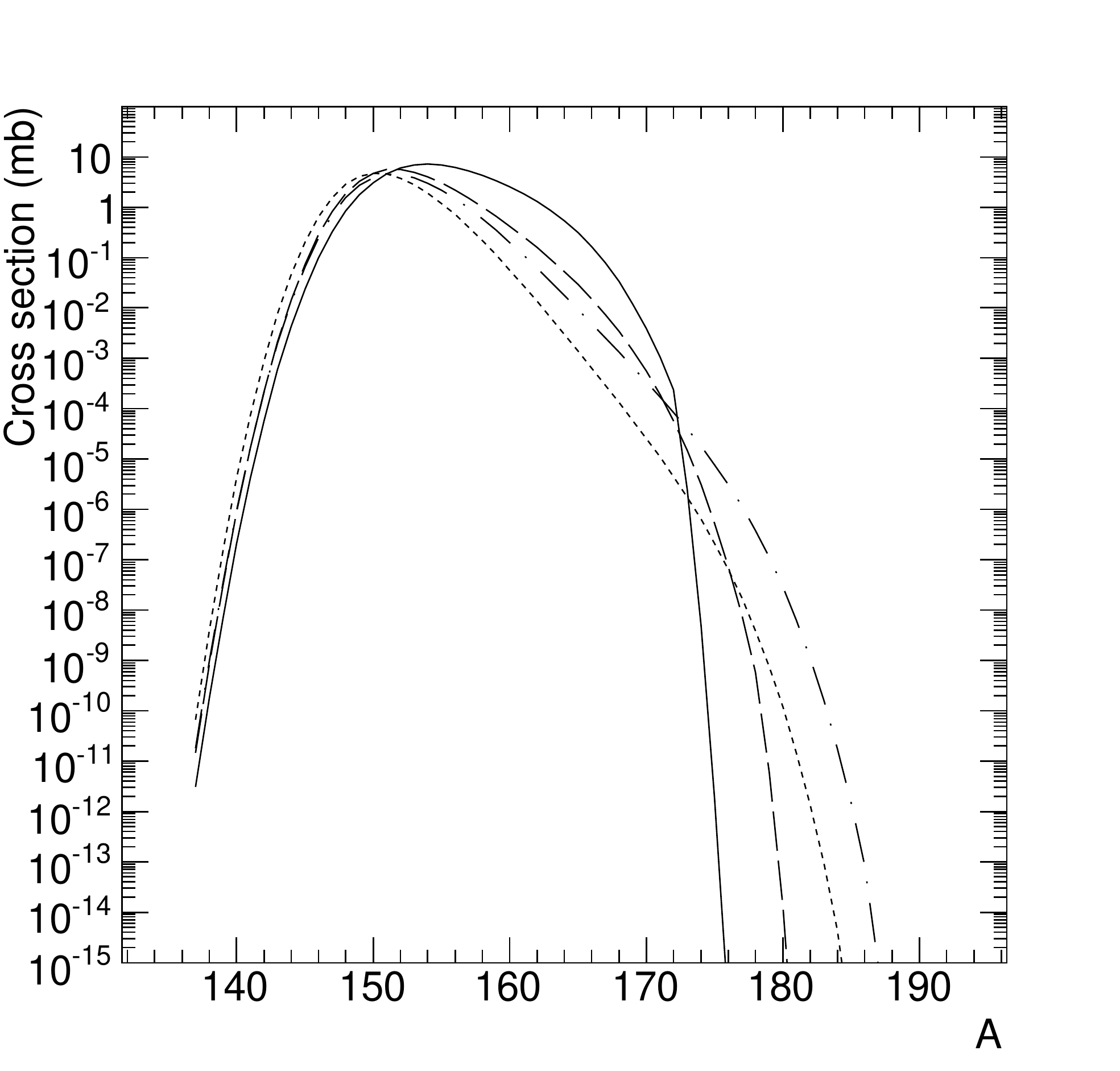}
 \caption[Grazing calculations of production cross-sections for dysprosium isotopes]{Grazing calculations of production cross-sections for dysprosium isotopes (left) using an \isotope{170}{Er} target and a \isotope{48}{Ca} beam with an energy of 230~MeV (dotted), a \isotope{82}{Se} beam with an energy of 460~MeV (dashed) \protect\cite{PhysRevC.81.034310} and a \isotope{136}{Xe} beam with an energy of 1000~MeV (solid).
Calculated production cross-sections of the Dy isotopic chain in the FRS (right) for the primary beams \isotope{176}{Yb} (solid), \isotope{186}{W} (dashed), \isotope{197}{Au} (dotted) and \isotope{198}{Pt} (dash-dotted). See \protect\cite{my_thesis} for details.}
 \label{fig:grazing-Dy}
\end{figure}

Taking advantage of new detector technology, like AGATA \cite{agata}, and the possibility to use a heavier beam, like \isotope{136}{Xe}, both the detection power and production cross-section of the experiment described in \cite{PhysRevC.81.034310} could be improved considerably. Thus, it would be possible to extend the systematic studies of collectivity further into the neutron-rich region using multi-nucleon transfer reactions. The production cross-section of dysprosium isotopes for three different types of ion beams calculated using the \texttt{grazing} code \cite{grazing1,grazing2} are shown in figure~\ref{fig:grazing-Dy}.

A valuable complement to the multi-nucleon transfer reaction measurements are the fragmentation reactions. Using these it is possible to establish $B(\mathrm{E2})$ values to determine the electric quadrupole moments and the degree of triaxial deformation, and thus the evolution of quadrupole collectivity, for a range of neutron-rich rare-earth nuclei \cite{0954-3899-36-11-115104}. By relativistic Coulomb excitation it would be possible to determine the \betwo{0^{+}}{2^{+}_{1}} and \betwo{0^{+}}{2^{+}_{2}} for a large range of neutron-rich even-even nuclei in the rare-earth region \cite{pregan}, see for example the Dy isotopes in figure~\ref{fig:grazing-Dy}. In this figure the production cross-sections, calculated by the LISE++ code \cite{lise1,lise2}, for a couple of primary beams from the fragment separator (FRS) at GSI with energies
of 800~MeV per nucleon and
projectile fragmentation on a 4~g/cm$^{2}$ beryllium target are shown.
A problem with this kind of measurement, however, is the large background from bremsstrahlung radiation for $\gamma$-ray energies $E_{\gamma}<300$~keV, why detailed simulations are required.

\end{document}